\documentclass{llncs}

\usepackage[english]{babel}
\usepackage[utf8x]{inputenc}
\usepackage[T1]{fontenc}

\usepackage{graphicx}
\usepackage[colorinlistoftodos]{todonotes}
\usepackage[colorlinks=true, allcolors=blue]{hyperref}

\usepackage{xcolor}
\usepackage{xspace}
\usepackage{multicol}
\usepackage[lofdepth,lotdepth,caption=false]{subfig}

\newcommand\TOOL{OASGraph\xspace}

\begin{document}

\title{Generating GraphQL-Wrappers for REST(-like) APIs}

\author{Erik Wittern\inst{1} 
        \and Alan Cha\inst{2}\thanks{Work performed during Summer Internship at IBM Research in 2017}
        \and Jim A. Laredo\inst{1}}

\institute{IBM T.J Watson Research Center, Yorktown Heights, NY, 10598 USA\\
\email{witternj@us.ibm.com, laredoj@us.ibm.com}
\and
Columbia University, New York, NY, 10025 USA \\
\email{ac3805@columbia.edu}}

\maketitle

\begin{abstract}
GraphQL is a query language and thereupon-based paradigm for implementing web Application Programming Interfaces (APIs) for client-server interactions. Using GraphQL, clients define precise, nested data-requirements in typed queries, which are resolved by servers against (possibly multiple) backend systems, like databases, object storages, or other APIs. Clients receive only the data they care about, in a single request.
However, providers of existing REST(-like) APIs need to implement additional GraphQL interfaces to enable these advantages.
We here assess the feasibility of automatically generating GraphQL wrappers for existing REST(-like) APIs. A wrapper, upon receiving GraphQL queries, translates them to requests against the target API.
We discuss the challenges for creating such wrappers, including dealing with data sanitation, authentication, or handling nested queries.
We furthermore present a prototypical implementation of \TOOL. \TOOL takes as input an OpenAPI Specification (OAS) describing an existing REST(-like) web API and generates a GraphQL wrapper for it.
We evaluate \TOOL by running it, as well as an existing open source alternative, against $959$ publicly available OAS. This experiment shows that \TOOL outperforms the existing alternative and is able to create a GraphQL wrapper for $89.5\%$ of the APIs -- however, with limitations in many cases. A subsequent analysis of errors and warnings produced by \TOOL shows that missing or ambiguous information in the assessed OAS hinders creating complete wrappers.
Finally, we present a use case of the IBM Watson Language Translator API that shows that small changes to an OAS allow \TOOL to generate more idiomatic and more expressive GraphQL wrappers.
\end{abstract}

\section{Introduction}
\label{sec:introduction}
Created by Facebook in 2012 and released as open source in 2015, \emph{GraphQL} is a query language and thereupon-based paradigm for building web Application Programming Interfaces (APIs) for client-server interactions. According to the lore\footnote{\url{https://code.facebook.com/posts/1691455094417024/graphql-a-data-query-language/}}, 
Facebook created GraphQL in response to observed issues with more conventional ways of building APIs, like the Representational State Transfer (REST) architectural style~\cite{Fielding:2000} or one of its variants (we refer to this style of APIs, broadly, as \emph{REST(-like)}, because many such APIs do not fully adhere to the constraints prescribed by REST~\cite{Rodriguez:2016}).
In REST(-like) APIs, resources are identified by URIs and are accessed or manipulated via (most commonly) HTTP endpoints.
As a result, clients are limited to perform predetermined operations, which may have been designed by API providers irrespective of the clients' specific requirements.
In consequence, clients frequently receive unneeded data, or have to chain multiple requests to obtain desired results.
Furthermore, for API clients, changes to a REST(-like) API can have sever implications, and in the worst case lead to application misbehavior or even crashes~\cite{Li:2013,Espinha:2014}.
Another issue is that REST(-like) APIs amass endpoints when providers add new capabilities to an API to preserve compatibility -- for example to react to new client requirement without breaking compatibility with existing clients. 

GraphQL's solution to these problems is a query language that allows clients to specify exact data requirements (within provider-defined constraints) on a data field level, thus avoiding to send superfluous data over the network. Changes to client-specific requirements can be resolved by changing queries, rather than having to add endpoints to the API. A graph-based data abstraction allows providers to add new capabilities to their data model without breaking client code.
(Arbitrarily) nested queries allow to combine previously multiple requests, reducing client code complexity and avoiding associated overhead.
Due to these advantages, various prominent API providers offer GraphQL interfaces (in addition to their REST(-like) APIs), among them GitHub\footnote{\url{https://developer.github.com/v4/}}, Yelp\footnote{\url{https://www.yelp.com/developers/graphql/guides/intro}}, and The New York Times\footnote{\href{https://open.nytimes.com/react-relay-and-graphql-under-the-hood-of-the-times-website-redesign-22fb62ea9764}{\texttt{https://open.nytimes.com/react-relay-and-graphql-under-the-hood-of-the-times-website-\\ redesign-22fb62ea9764}}}.

To offer a GraphQL interface, providers have to implement, operate and evolve it -- possibly in addition to existing REST(-like) APIs. To delimit this burden on providers, in this work, we assess the feasibility of leveraging existing REST(-like) APIs and their machine-readable \emph{specifications} to automatically generate a GraphQL \emph{wrapper}. The wrapper resolves GraphQL queries by performing requests against the existing API. To the best of our knowledge, this is the first scientific work addressing the generation of GraphQL wrappers for REST(-like) APIs.

In the remainder of this work, we start in Section~\ref{sec:background} by providing a more detailed description of GraphQL and REST(-like) APIs definitions, specifically the OpenAPI Specification. In Section~\ref{sec:translation}, we discuss the challenges of generating GraphQL wrappers for existing REST(-like) APIs, and possible solutions. In Section~\ref{sec:implementation}, we present a proof-of-concept implementation, \emph{\TOOL}, that, for a given OpenAPI Specification, generates a GraphQL wrapper. In Section~\ref{sec:evaluation}, we evaluate \TOOL and an existing open source tool by applying them to $959$ public  OpenAPI Specifications. We assess the errors produced by both tools, and warnings produced by \TOOL, which indicate partial output generation. The evaluation reveals that \TOOL improves upon the state of the art in generating GraphQL wrappers, and that encountered issues are largely to blame on missing or ambiguous information in the assessed OpenAPI Specifications. In light of these results, we present a use case in Section~\ref{sec:usecase} for a well-specified API and enhance it with small changes to render the resulting GraphQL wrapper more idiomatic and expressive. We present related work in Section~\ref{sec:related_work} before discussing our work and concluding in Section~\ref{sec:conclusion}.

\section{Background}
\label{sec:background}
In this Section, we discuss GraphQL as well as on OpenAPI Specifications, which act as input for creating GraphQL wrappers for existing REST(-like) APIs.

\subsection{GraphQL}
\label{sec:background_graphql}
GraphQL describes itself as a query language ``[...] for describing the capabilities and requirements of data models for client-server applications.''~\cite{GraphQLSpec}. In a condensed form, the interactions of using GraphQL comprise of the following:

\begin{itemize}
  \item A GraphQL server implements a \emph{GraphQL schema} that defines the \emph{types} and relations of exposed data, including \emph{operations} to query or mutate data. A schema can define a \texttt{user} data as an object that contains string fields \texttt{id} and \texttt{name} and an object field \texttt{address}, with string fields \texttt{street} and \texttt{city}. Clients can query users by providing a string \texttt{id} argument. Data types can be associated with \emph{resolve functions}, which implement operations against arbitrary backend systems such as a database.
  \item A GraphQL client \emph{introspects} a server's schema, i.e., queries it with GraphQL queries, to learn about exposed data types and possible operations. The \emph{GraphiQL}\footnote{\url{https://github.com/graphql/graphiql}} online-IDE uses introspection to allow developers to familiarize with GraphQL schemas.
  \item A client sends \emph{queries} to the server, whose syntax resembles that of the JavaScript Object Notation (JSON) and which specify desired operations to perform and what data to return (on the level of fields of objects).
  \item Upon receiving queries, the server \emph{validates} them against the schema and \emph{executes} them by invoking one or more \emph{resolve functions} to either fetch the requested data or perform desired mutations.
  \item Ultimately, the server sends back requested data to the client, or error messages in case the execution failed.
\end{itemize}

\subsection{OpenAPI Specifications}
\label{sec:background_oas}
The \emph{Open API Specification} (OAS), formerly known as Swagger, is ``a standard, programming language-agnostic interface description for REST APIs''~\cite{Oas3}. OAS is a format used by providers to describe and document APIs in an organized and predictable manner. Both human and machine-based clients use OAS to understand and invoke APIs, including tooling that works for any API described using OAS. OAS breaks an API down into operations identified by a unique combination of URL path and HTTP method, the data in- and output schemas (both for successful responses as well as errors), required parameters (e.g., in headers or query strings), and authentication mechanisms.

\section{Generating a GraphQL Schema from an OAS}
\label{sec:translation}
Within this section, we describe how to create a GraphQL wrapper for a target REST(-like) API discussing the challenges and proposing ways to mitigate them. In general, the presented approach relies on taking as input an OAS describing the target API and outputting a GraphQL schema that, once deployed, forms a GraphQL wrapper around the target API. The GraphQL wrapper translates queries to corresponding requests against the target API. The schema consists of
a) the data types expected and exposed by the wrapper and their relations (see Section~\ref{sec:translation_data}) and
b) resolve functions responsible for receiving and returning data by making requests to the target API (see Section~\ref{sec:translation_resolve}).
We also discuss the support nested queries (see Section~\ref{sec:translation_links}) and handling authentication requirements of target APIs (see Section~\ref{sec:translation_auth}), before describing how all these pieces are ultimately combined to form a GraphQL schema (see Section~\ref{sec:translation_schema}).

\subsection{Translating Schema Objects to GraphQL Types}
\label{sec:translation_data}
A GraphQL schema uses \emph{GraphQL types} to define the data being sent to or returned from a GraphQL interface~\cite{GraphQLSpec}. Most notably, \emph{(Input) Object types} define the structure of JSON objects.\footnote{GraphQL is agnostic to the used data serialization format. However, due to its predominance and first-class support in OAS, we assume and speak of JSON as the serialization format throughout this paper.} (Input) Object types contain named fields whose values can either be other (Input) Object types, \emph{List types}, \emph{Enum types}, or \emph{Scalar types} (like Int, Float, String, or Boolean). List types contain items of any other type, while Enum types define allowed values of type String.

To define the GraphQL types of a target API, one can make use of the \emph{schema objects} defined in that API's OAS (not to be confused with the GraphQL schema). Schema objects largely comply with the \emph{JSON Schema Specification}~\cite{JSONSchema}, a format used to describe the structure of JSON data. Often, schema objects can be directly mapped to GraphQL types. For instance, JSON Schema \texttt{object}s map to GraphQL (Input) Object types, \texttt{array}s map to GraphQL List types, and \texttt{enum}s map to GraphQL Enum types. Similarly, scalar types in JSON Schema like \texttt{string}, \texttt{boolean}, or \texttt{number} / \texttt{integer} match to corresponding GraphQL Scalar types. In consequence, our approach is to iterate through the OAS' schema objects and instantiate corresponding GraphQL types for each one of them.

\subsubsection{Schema Object De-Duplication and Type Naming}
\label{sec:translation_data_dedup}
One challenge in defining GraphQL types is to avoid duplicate (Input) Objects. (Input) Object types in a single GraphQL schema need to have unique names as identifiers. Furthermore, duplicate types, with different names, lead to bloated GraphQL schemas and possibly cause users confusion. However, the OpenAPI Specification, while providing a \emph{reference} mechanisms to foster reuse of schema objects defined in a central \texttt{components} object, does not enforce their de-duplication.

Schema object de-duplication can be achieved by creating a \emph{types dictionary} for a given OAS in a pre-processing phase (i.e., prior to generating GraphQL types). The dictionary contains all schema objects defined across all operations in an OAS. A new schema object is only added if a deep comparison attests it to be unique. The types dictionary further flattens out nested schema objects. That is, if a schema defines an object with properties that are themselves objects, dedicated entries for the latter are created in the types dictionary.

A unique name string is required to identify types in a GraphQL schema, to associate types with the operations that consume/produce data of that type, and to identify entries in the types dictionary. From a given OAS, names can be derived from an explicit reference to the schema object if it appears in the components object of an OAS (e.g., \texttt{User} from \texttt{"\#/components/schema/User"}), or from an explicitly set \texttt{title} value in the schema object (if present). If a schema object is not referenced and does not have a title, or if any of these values has already been used in the types dictionary, the following fall-backs can be used: if the schema object was referenced from an operation, a concatenation of this operations HTTP method and URL path is used. If, however, the schema object stems from the definition of another, complex schema object, the key identifying the schema object in that context is used.

\subsubsection{Translation Process}
\label{sec:translation_data_translation}
Once a types dictionary has been created, every contained schema object can be translated to a GraphQL type. The translation approach depends on the \texttt{type} property of a schema object:

\begin{itemize}
  \item If the schema object defines a scalar type (e.g., \texttt{string}, \texttt{number}, or \texttt{boolean}), a corresponding GraphQL Scalar Type is created. As an exception, if the type is \texttt{string} but valid \texttt{enum} values are defined, the translation creates a corresponding GraphQL Enum Type.

  \item If the schema object defines an \texttt{object}, all its properties (in JSON Schema terms) are traversed and corresponding fields are added to a new (Input) Object Type. Here, an \emph{Input} Object Type is created if the schema object defines the payload of any operation, and a normal Object Type is created if the schema object defines the response data of any operation.\footnote{GraphQL requires to use dedicated Input Object types to define objects sent as arguments in queries because they delimit otherwise available mechanisms for polymorphism like \emph{Union Types} or \emph{Interfaces}, which may not be unambiguously cast to their correct type.} The values of the created fields are themselves GraphQL types, reflecting again the JSON schema property types. In consequence, a recursive algorithm is required to translate possibly nested objects and to create equally nested (Input) Object Types. Once created, (Input) Object Types are stored in the types dictionary for possible reuse. In cases where an object's property is itself of type \texttt{object}, the corresponding type can then either be referenced (if it had already been translated), or the translation of that type is triggered.
    
    \item If the schema object defines an \texttt{array}, first, a new GraphQL type describing the array's items needs to be created. Once a GraphQL type defining the items has been created, it is wrapped in a GraphQL List Type.
\end{itemize}

In all of the above cases, properties marked as \texttt{required} in a schema object are wrapped in GraphQL \emph{NonNull types} to express the same requirement. Key for Input Object types, as it forces users to provide all data required by the target API.
Any schema object with human-readable \texttt{description} is exposed in the created GraphQL types, and made available to developers during introspection.

\subsubsection{Data Sanitation}
\label{sec:translation_data_sanitation}
Names of types, arguments, and (Input) Object type fields need to follow the GraphQL specification. It requires them to adhere to the regular expression \texttt{/[\_A-Za-z][\_0-9A-Za-z]*/}, i.e., must start with underscore (``\_'') or a letter, and may then contain only underscores and alphanumerics~\cite{GraphQLSpec}. Because similar restrictions do not exist for REST(-like) APIs defined in an OAS, sanitation requires to remove any non-supported characters. Sanitation, however, has multiple effects:
First, it causes the GraphQL wrapper to deviate from the REST(-like) API. 
Second, the resolve functions of the GraphQL wrapper need to un-sanitize data it sends to the target API, and sanitize responses for them to match with the sanitized GraphQL schema. For this purpose, a mapping between raw and sanitized values needs to be built up during type creation and made accessible to the resolve functions (see Section~\ref{sec:translation_resolve}).

\subsection{Creating Resolve Functions}
\label{sec:translation_resolve}
Resolve functions make requests to the target API in response to GraphQL queries, to either retrieve or mutate data. Resolve functions can be created and returned by a ``generator'' function for a given operation defined in an OAS.

The generator binds information from an OAS needed by the resolve function to perform requests \emph{during their creation}. The generator binds the operation's \emph{baseUrl}, \emph{URL path}, \emph{HTTP method}, and information about supported authentication mechanisms. Furthermore, the generator binds a mapping between the names of arguments that a resolve function may receive as input from a query and the instructions for sending those arguments as \emph{parameters} of a request to the target API. In REST(-like) APIs, identifiers of resources and other smaller pieces of data is typically sent as \emph{path parameters} or \emph{query parameters} or in \emph{headers}. More complex data is typically sent as \emph{payload} in the request body of, for example, \texttt{POST}, \texttt{PUT}, or \texttt{PATCH} requests. Finally, resolve functions should be aware of  \emph{default} values, which, if defined, are used if a query does not provide a value for an argument. All this information can be found as part of the OAS.

Resolve functions should be able to receive and process other pieces of information at \emph{runtime}, including data received from previous, parent resolve functions, this being a default behavior of GraphQL.
Resolve functions may also receive security-related information like API keys, credentials, or OAuth / OpenID Connect tokens via a \emph{context} object that is available across resolve functions, another default behavior of GraphQL.
Finally, arguments used in previous requests are also passed down at runtime, so that they only have to be defined once per query, even if used by multiple resolve functions.

A previously created mapping between raw and sanitized values, as described in Section~\ref{sec:translation_data_sanitation}, is used by resolve functions in two ways:
Before sending a request, passed argument names are de-sanitized. For example, the received argument \texttt{\{"id": 1\}} may be de-sanitized to \texttt{\{"\$id": 1\}}, as the target API expects a payload in the latter form.
After receiving a response from the target API, resolve functions sanitize received data for it to be properly handled by the GraphQL runtime.
(De-) sanitation is of a recursive nature as it covers nested objects and arrays to assign the requested returned values.

\subsection{Nested Data via ``Links''}
\label{sec:translation_links}
\begin{figure}[ht]
\centering
\subfloat[Subfigure 1 list of figures text][OAS link definition in YAML.]{
  \includegraphics[width=0.45\textwidth]{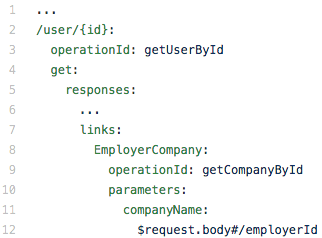}
  \label{fig:link-example}
}
\qquad
\subfloat[Subfigure 2 list of figures text][GraphQL query using link.]{
  \includegraphics[width=0.25\textwidth]{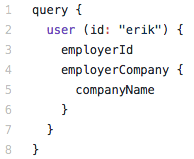}
  \label{fig:link-query-example}
}
\caption{Link examples.}
\end{figure}

A distinguishing feature of GraphQL is querying (deeply) nested data in single request. In REST(-like) APIs, similar operations may involve multiple requests. As of version 3.0.0, introduced in June 2017, an OAS can define possible combinations of requests using \texttt{links}. A link provides design-time information\footnote{In that sense, links differ from hypermedia provided by RESTful APIs during \emph{runtime} as part of Hypermedia as the Engine of Application State (HATEOAS) constraint~\cite{Fielding:2000}.} about the relationships between the response of a requests and possibly subsequent requests, which depend on this response. For example, Figure~\ref{fig:link-example} shows a link definition of an OAS written in YAML, which states that the \texttt{employerId} returned in the payload when invoking \texttt{GET ../user/\{id\}} can be used to instantiate the \texttt{companyName} parameter in a request to the \texttt{getCompanyById} operation.

A link defined in an OAS operation may create an additional field in the operation's response GraphQL Object type. The name of that field is the (sanitized) identifier of the link (e.g., \texttt{EmployerCompany}) and the type of that field is the type of the response data of the linked operation. In the example query in Figure~\ref{fig:link-query-example}, a client fetches data on a user with id ``erik'' and uses the field \texttt{employerCompany}, created based on the link, to also fetch the employer's \texttt{companyName}.

For such queries to work, resolve functions need to be able to receive parameters from previous, parent resolve functions. For example, to resolve the GraphQL query in Figure~\ref{fig:link-query-example}, one resolve function invokes the \texttt{getUserById} operation and passes the received \texttt{employerId} field to a second resolve function that uses this data to invoke the \texttt{getCompanyById} operation.

\subsection{Authentication}
\label{sec:translation_auth}
A GraphQL wrapper needs to implement required API authentication mechanisms. We here present ways to support a) API key and basic authentication and b) OAuth 2 or OpenID Connect in GraphQL wrappers.

\subsubsection{API Keys and Basic Authentication}
\label{sec:translation_auth_basic}
\emph{Authentication viewers} provide a mechanism to allow users to pass authentication information (API keys, or username and password). Viewers are special GraphQL (Input) Object types that wrap all other GraphQL (Input) Object types whose resolve functions require authentication. Viewers define mandatory arguments \texttt{apiKey} or \texttt{username} and \texttt{password} and propagate their values to the resolve functions of all wrapped child types (see Section~\ref{sec:translation_resolve}).
Viewers place sensitive credentials in GraphQL queries, requiring dedicated security mechanisms (i.e., transport encryption).

\subsubsection{OAuth 2 and OpenID Connect}
\label{sec:translation_auth_oauth}
\emph{OAuth 2} is an authorization framework where users rely on a third party service's \emph{OAUTH server} to authenticate and authorize certain actions of an application. Applications are registered with the OAUTH server, and subsequently forward users to the server. Users authenticate themselves with the OAUTH server, which returns \emph{access tokens} to the application. The application uses the tokens to interact with protected resources.

\emph{OpenID Connect} is a layer on top of OAuth 2 prescribing the use of JSON Web Token, equivalent to OAuth 2 for the purpose of a GraphQL wrapper.

The flow outlined above is independent of a GraphQL wrapper itself, rather, it relies on the application (e.g., the server hosting the GraphQL wrapper) to obtain the necessary tokens. The resolve functions of a GraphQL wrapper, need to be able to a) obtain access tokens from an application and b) send these tokens within requests to a target REST(-like) API, typically by including them in an \texttt{Authorization} header. We describe one way of passing tokens to a GraphQL wrapper when presenting our proof-of-concept implementation in Section~\ref{sec:implementation}.

\subsection{Building up the GraphQL Schema}
\label{sec:translation_schema}
Having translated schema objects to GraphQL types (relying on a types dictionary; possibly considering links) and having defined resolve functions for every operation, an overall GraphQL schema can be created, defining all possible queries and mutations.
If authentication viewers were created, they are added as root elements to the queries and mutations fields.
Then, for every operation defined in the OAS, the created response type, input types (forming arguments), and resolve function are collectively added to the query or mutation fields of the schema, depending on whether the HTTP method of the operation is \texttt{GET} or not, either directly, or within the previously added authentication viewers they depend on.
The resulting GraphQL schema can be passed to a GraphQL server implementation and deployed to start receiving queries.

\section{Implementation}
\label{sec:implementation}
We created a proof-of-concept called \TOOL implementing the concepts for automatically wrapping REST(-like) APIs with GraphQL as described in Section~\ref{sec:translation}.
\TOOL is written in JavaScript using the Flow static type checker\footnote{\url{https://flow.org}}, and relying on the GraphQL reference implementation GraphQL.js.\footnote{\url{http://graphql.org/graphql-js}}
\TOOL further relies on a third party library \emph{swagger2openapi}\footnote{\url{https://github.com/Mermade/swagger2openapi}} to translate given OAS 2.0 (Swagger) specifications to OAS 3.0.0\footnote{Because version 3.0.0 is a superset of Swagger, no information is lost during this translation.}, and to validate that provided specifications are syntactically correct.

As proposed in Section~\ref{sec:translation}, \TOOL performs a pre-processing phase to de-duplicate and name schema objects (see Section~\ref{sec:translation_data_dedup}), before recursively translating them to GraphQL types (see Section~\ref{sec:translation_data_translation}) while sanitizing type, argument, and field names (see Section~\ref{sec:translation_data_sanitation}). During the translation, \TOOL resolves possibly encountered references within the given OAS (e.g., \texttt{\$ref: "\#/components\\/schemas/User"}) as well as \texttt{allOf} definitions\footnote{For details, see \url{https://github.com/OAI/OpenAPI-Specification/blob/master/versions/3.0.1.md\#schema-object}}, and enriches Object types with fields stemming from link definitions (see Section~\ref{sec:translation_links}). \TOOL further generates authentication viewers for passing API keys and basic authentication credentials (see Section~\ref{sec:translation_auth}), as well as an \emph{any auth viewer} that takes as arguments multiple authentication information at once, allowing nested queries to rely on more than one authentication mechanism.
After creating resolve functions per operation in the given OAS (see Section~\ref{sec:translation_resolve}), \TOOL combines them with the generated types and authentication viewers to form a GraphQL schema (see Section~\ref{sec:translation_schema}).

\TOOL further allows to pass a \emph{JsonPath}~\cite{JsonPath} option that points resolve functions to the location of authentication tokens in the global context object (see Section~\ref{sec:translation_resolve}). This option is used to provide resolve functions access to OAuth 2 or OpenID Connect tokens made available in the context by the application deploying the GraphQL interface.

Finally, \TOOL provides two modes of operation:
In \emph{strict} mode, \TOOL will throw errors in light of missing or ambiguous information in a given OAS. Strict mode aims to create a GraphQL wrapper that is complete and closely aligned with the target API, or no wrapper at all if that is impossible.
In contrast, in \emph{non-strict} mode, \TOOL attempts to mitigate lacking or ambiguous information in a given OAS, leading to a working GraphQL wrapper, that may slightly deviate from the target API. In non-strict mode, \TOOL tracks causes for such cases and performed mitigations as \emph{warnings} in a report, which is made accessible to applications or developers. We present the types of warnings (as well as their occurrences) and the mitigations performed by \TOOL as part of the quantitative evaluation in Section~\ref{sec:evaluation_results_warnings}.

The GraphQL schema produced by \TOOL can be used by any GraphQL-compliant JavaScript framework, like the \emph{express-graphql}\footnote{\url{https://github.com/graphql/express-graphql}} library to run the GraphQL wrapper as an Express.js application.

\section{Quantitative Evaluation}
\label{sec:evaluation}
The goal of this section is to investigate the feasibility of automatically wrapping \emph{any} REST-like API with GraphQL. To answer this question, we applied \TOOL as well as \emph{Swagger2GraphQL}, an open source tool with the same goal, to a large number of publicly available OAS. For both tools, we assess the causes of errors that occurred during the experiments. In addition, for \TOOL, we assess warnings produced in non-strict mode to analyze to what degree the created GraphQL wrappers cover the target REST(-like) APIs.

\subsection{Data Collection}
\label{sec:evaluation_data}
For the evaluation, we obtained $959$ OAS made available in the \emph{APIs.guru OpenAPI Directory}\footnote{\url{https://apis.guru/openapi-directory}}. In this directory, third parties maintain OAS 2.0 of popular APIs. These OAS are created by dedicated scripts that either translate other API specification formats to OAS, or extract required information from (human-readable) API documentations (typically written in HTML). APIs.guru runs these scripts weekly, and manually checks detected differences for correctness before committing them. In addition, error-fixes can be contributed by a larger community through pull requests on the directory's GitHub repository\footnote{\url{https://github.com/APIs-guru/openapi-directory}}. We collected the evaluation data on January 11th 2018.

\subsection{Experiment Execution}
\label{sec:evaluation_experiments}
We ran \TOOL, once in strict and once in non-strict mode, on all $959$ OAS contained in the APIs.guru dataset.
We repeated the experiment with Swagger2GraphQL, an existing open source tool that, in the same way as \TOOL, aims to automatically generate GraphQL wrappers for existing REST(-like) APIs.
Like \TOOL, Swagger2GraphQL iterates through the operations defined in a given OAS and creates corresponding GraphQL types, including mutations for non-\texttt{GET} operations.
In contrast to \TOOL, Swagger2GraphQL relies on OAS in version 2.0 (``Swagger'', hence the eponymous name) as input, meaning it does not consider links for nested API requests (links do not exist in OAS 2.0).
Swagger2GraphQL does not de-duplicate schemas objects,
does not sanitize type, argument, and field names,
does not consider enum types,
and does not provide viewer types for basic authentication and API keys.
Furthermore, Swagger2GraphQL has a different approach to dealing with the risk of duplicate GraphQL type names:
rather than inferring unique names for types (see Section~\ref{sec:translation_data_dedup}), Swagger2GraphQL relies on the \texttt{operationId} provided in a given OAS or falls back to combining the HTTP method and path of an operation. In consequence, query types have names like ``getUserById'', which are unique but arguably less idiomatic for GraphQL.

\subsection{Results}
\label{sec:evaluation_results}
We evaluate the results of the quantitative evaluation in terms of the \emph{errors} and \emph{warnings} produced by \TOOL (and Swagger2GraphQL) during our experiments.

\subsubsection{Errors}
\label{sec:evaluation_results_errors}
Table~\ref{table:overall_results} summarizes the number of cases with and without errors when applying \TOOL (both in strict and non-strict mode) and Swagger2GraphQL to the $959$ OAS from the APIs.guru dataset.

\begin{table}[ht]
\begin{center}
  \begin{tabular}{l | c c }
           & \textbf{Succ. (\%)} & \textbf{Errors (\%)} \\
    \hline
    OASGraph (non-strict) & 930   (97\%) &  29    (3\%) \\
    OASGraph (strict)     & 260 (27.1\%) & 699 (72.9\%) \\    
\hline
    Swagger2GraphQL       & 591 (61.6\%) & 368 (38.4\%)
\end{tabular}
\end{center}
\caption{Overall Results}
\label{table:overall_results}
\end{table}

As can be seen from Table~\ref{table:overall_results}, \TOOL produces significantly more errors in strict mode, in which case only just over a quarter of cases succeeds. On the other hand, in non-strict mode, wrapping an API succeeds in over $95\%$ of cases. The difference in these values motivates a detailed investigation into the mitigations performed by \TOOL and thus the deviation of created GraphQL interfaces from their target APIs.

The result of Swagger2GraphQL, on first sight, falls between those of \TOOL in strict and non-strict mode. Swagger2GraphQL performs better than \TOOL in strict mode because it (silently) mitigates issues with the input OAS that cause \TOOL to fail in strict mode. For example, Swagger2GraphQL silently creates a ``dummy'' GraphQL Object type if an operation does not define a valid response schema object, or it arbitrarily selects one HTTP status code for which to define a response type in case multiple codes are available. In both cases, \TOOL throws an error in strict mode (and mitigates these issues with a warning in non-strict mode). As such, the number of APIs for which Swagger2GraphQL succeeds to create a GraphQL wrapper would be smaller, if these cases were explicitly exposed.
On the other hand, Swagger2GraphQL performs worse than \TOOL in non-strict mode, as it, for example, does neither sanitize field, argument, and type names nor de-duplicate type names.

\begin{table}[ht]
\centering
\begin{tabular}{l | c c }
  \textbf{Error type} & \textbf{OASGraph (non-strict)} \quad & \textbf{Swagger2GraphQL} \\ \hline
  Invalid OAS          &  6 &   0 \\
  Sanitation Error     & 16 &   0 \\
  Missing Ref          &  7 &   7 \\
  Name Conflict        &  0 &   2 \\
  Unknown Schema Type  &  0 &  35 \\
  No Get Operation     &  0 &  20 \\
  Unsanitized Name     &  0 & 252 \\
  Invalid Schema Type  &  0 &  25 \\
  Stack Overflow       &  0 &  27 \\
  \hline
  Overall              & 29 & 368
\end{tabular}
\caption{Breakdown of errors produced by \TOOL (in non-strict mode) and Swagger2GraphQL.}
\label{table:errors}
\end{table}

Table~\ref{table:errors} breaks down the types of errors produced by \TOOL in non-strict mode\footnote{We do not cover errors thrown by \TOOL in strict mode here, because they are captured as warnings in non-strict mode, which we break down later in this section.} and by Swagger2GraphQL. The explanation for the error codes in Table~\ref{table:errors} is as follows:

\begin{itemize}
  \item \textbf{Invalid OAS}: The input OAS could not be successfully validated (by the third party library swagger2openapi used by \TOOL). Because Swagger2GraphQL does not perform validation, it does not produce such errors. 

  \item \textbf{Sanitation Error}: Sanitation of type, argument, or field names fails (see Section~\ref{sec:translation_data_sanitation}). The errors thrown by \TOOL result from attempts to sanitize enum values of type boolean. While such enumeration values are valid in schemas objects, they are not valid in GraphQL enum types, specifically, the strings \texttt{true} and \texttt{false} are forbidden as enum values~\cite{GraphQLSpec}. Because Swagger2GraphQL does not perform sanitation, it does not produce such errors. 

  \item \textbf{Missing Ref}: A reference cannot be resolved because it refers to relative documents\footnote{For details, see \url{https://github.com/OAI/OpenAPI-Specification/blob/master/versions/3.0.0.md\#relative-schema-document-example}} which are not provided by APIs.guru.

  \item \textbf{Name Conflict}: GraphQL.js throws an error because multiple types share the same name. Because \TOOL ensures unique names, it produces no such errors.

  \item \textbf{Unknown Schema Type}: A schema object defines a type that does not match any (scalar) GraphQL type (e.g., \texttt{undefined} or \texttt{file}). In \TOOL, such cases produce a warning rather than an error in non-strict mode, and the type is assumed to be string as a mitigation.

  \item \textbf{No Get Operation}: The given OAS does not contain any \texttt{GET} endpoints. \TOOL does not produce an error in such cases because it falls back to defining an empty root query type if no \texttt{GET} operation is present.

  \item \textbf{Unsanitized Name}: GraphQL.js throws an error due to unsanitized type or field names. Because \TOOL performs sanitation, it produces no such errors.

  \item \textbf{Invalid Schema Type}: A given schema object cannot be translated to a GraphQL type. We are uncertain about the origin of these errors in Swagger2GraphQL.

  \item \textbf{Stack Overflow}: JavaScript's maximum call stack is exceeded. This error seems to be a bug in Swagger2GraphQL's implementation.
\end{itemize}

As can be seen, by far the most errors produced by Swagger2GraphQL are caused by GraphQL.js throwing errors due to invalid type, argument, or field names. This finding underlines the importance of name sanitation as discussed in Section~\ref{sec:translation_data_sanitation}.

\subsubsection{Warnings}
\label{sec:evaluation_results_warnings}
The stark difference of results from running \TOOL in strict vs. non-strict mode motivates a detailed look into the types of mitigations \TOOL performs in non-strict mode. Overall, \TOOL reports $10,673$ warnings across all $930$ APIs that a GraphQL wrapper could be created for (called ``wrappable`` in the following). Of these, $260$ APIs could be wrapped without any warning (i.e., in strict mode) and the other $670$ APIs could be wrapped with at least one warning. The produced warnings are of the following types:

\begin{itemize}
  \item \textbf{5178 Missing Response Schema warnings}: An operation in the input OAS lacks a definition of a response or payload schema object. \TOOL's workaround is to ignore the operation. Swagger2GraphQL silently swallows such cases by generating a dummy ``empty'' default field of type string.

  \item \textbf{2502 Multiple Responses warnings}: An operation defines more than one response where the HTTP status code indicates success (i.e., is between 200 and 299). \TOOL's workaround is to select the lowest HTTP status code. Note that Swagger2GraphQL silently swallows such cases by randomly selecting the last status code between 200 and 299 defined in the given OAS.

  \item \textbf{2950 Invalid Schema Type warnings}: A payload or response schema object defined in the given OAS lacks a \texttt{type}, or is incomplete (for example, the \texttt{properties} definition of an \texttt{object} is empty). \TOOL's workaround is to fall back to assuming the type to be string. In consequence, clients can still receive such data as stringified JSON.

  \item \textbf{43 Unknown Schema Type warnings}: A schema object has a \texttt{type}, but that type is unknown to \TOOL (i.e., not \texttt{object}, \texttt{array}, \texttt{string}, \texttt{number}, or \texttt{boolean}). \TOOL's workaround, again, is to fall back to assuming the type to be string.
\end{itemize}



Looking into the distribution of warnings across APIs, we found that the majority of wrappable APIs have either no or few warnings of any particular type. Specifically, for every type of warning, half of the wrappable APIs have at most one warning of said type. While warnings are relatively concentrated to certain APIs when considered in isolation, they are less so when considered collectively: half of the wrappable APIs have over four warnings, and around one quarter of APIs have over 10 warnings of any type. In other words, it is not true that warnings overall are concentrated to few APIs, but rather that different APIs tend to have different warnings. This conclusion aligns with the previous observation that only $28\%$ ($260$) of wrappable APIs produce no warning at all.

Depending on the type of warning, \TOOL's mitigation strategies impact how complete the generated GraphQL wrappers are. We skip an operation on Missing Response Schema warnings to ensure that we have a fully usable wrapper rather than assume return codes and types and create a complete wrapper that will not behave properly.
For nearly half ($459$) of the wrappable APIs, all operations are translated.
About a quarter ($255$) of wrappable APIs skip under 25\% of operations, while around 12\% ($113$) skip 50\% or more of operations. Finally, 7.7\% (72) of wrappable APIs skip \emph{all} operations -- i.e., the resulting GraphQL wrappers are completely unusable and should be counted in addition to the $29$ APIs that \TOOL threw errors for even in non-strict mode.

\subsubsection{Discussion}
The quantitative evaluation reveals how challenging it is to wrap REST(-like) APIs automatically with GraphQL, given the state of existing, machine-readable API specifications. Ultimately, of the $959$ assessed APIs, only around a quarter ($260$) could be completely translated without warnings. In contrast, for 10.5\% ($101$) of APIs, translation completely failed ($29$ resulted in an error thrown by \TOOL, and for $72$ \TOOL could not wrap a single operation due to missing response schemas objects in the OAS). For the remaining $89.5\%$ of APIs ($858$), the completeness of the automatically generated GraphQL wrapper varies depending on the quality of the given OAS.

\section{Use Case: A GraphQL Wrapper for the IBM Watson Language Translator API}
\label{sec:usecase}
We present a use case showcasing how \TOOL can be applied for a well-specified API. Our use case centers around the \emph{IBM Watson Language Translator API}\footnote{\url{https://www.ibm.com/watson/services/language-translator}}. The API's main features include identifying the language of a given text (by \texttt{POST}ing the text to \texttt{.../v2/identify}), and translating a text between a number of languages (by \texttt{POST}ing the text to \texttt{.../v2/translate}). The public OAS\footnote{\url{https://watson-api-explorer.mybluemix.net/listings/language-translator-v2.json}} of the API is of version 2.0. \TOOL created a GraphQL wrapper for this API without warnings, as its OAS is complete and unambiguous.

\begin{figure}[ht]
\centering
\subfloat[Subfigure 1 list of figures text][Annotation to response body.]{
  \includegraphics[width=0.35\textwidth]{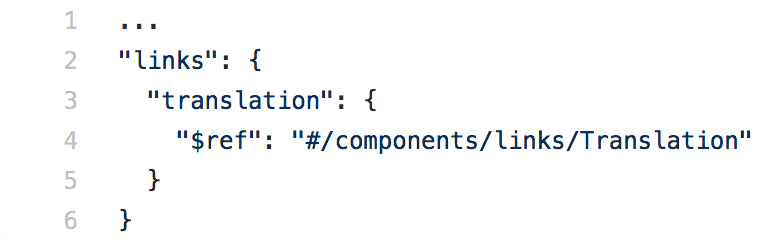}
  \label{fig:link_annotation}
}
\qquad
\subfloat[Subfigure 2 list of figures text][Prescriptive details to extract the linked elements.]{
  \includegraphics[width=0.5\textwidth]{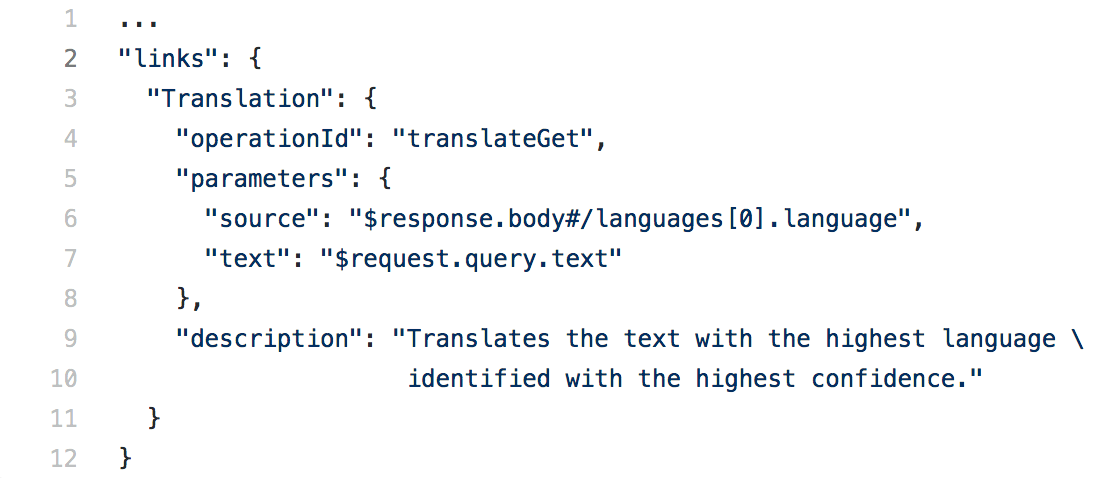}
  \label{fig:link_definition}
}
\caption{Link definition in the Watson Language Translator OAS.}
\end{figure}

To improve the GraphQL wrapper we loss-lessly translated it to version 3.0.0 using an online tool \footnote{\url{https://mermade.org.uk/openapi-converter}}.  Figure~\ref{fig:link_annotation} shows a \texttt{link} definition we added, and Figure~\ref{fig:link_definition} shows the definition of operations that have access to this link. In this case, the \texttt{translate} operation (labeled as \texttt{translateGet}) can use the first entry of the list of identified languages (thus, the most likely language of a text), and use it to define the required \texttt{source} argument.

\begin{figure}[ht]
\centering
\includegraphics[width=0.7\textwidth]{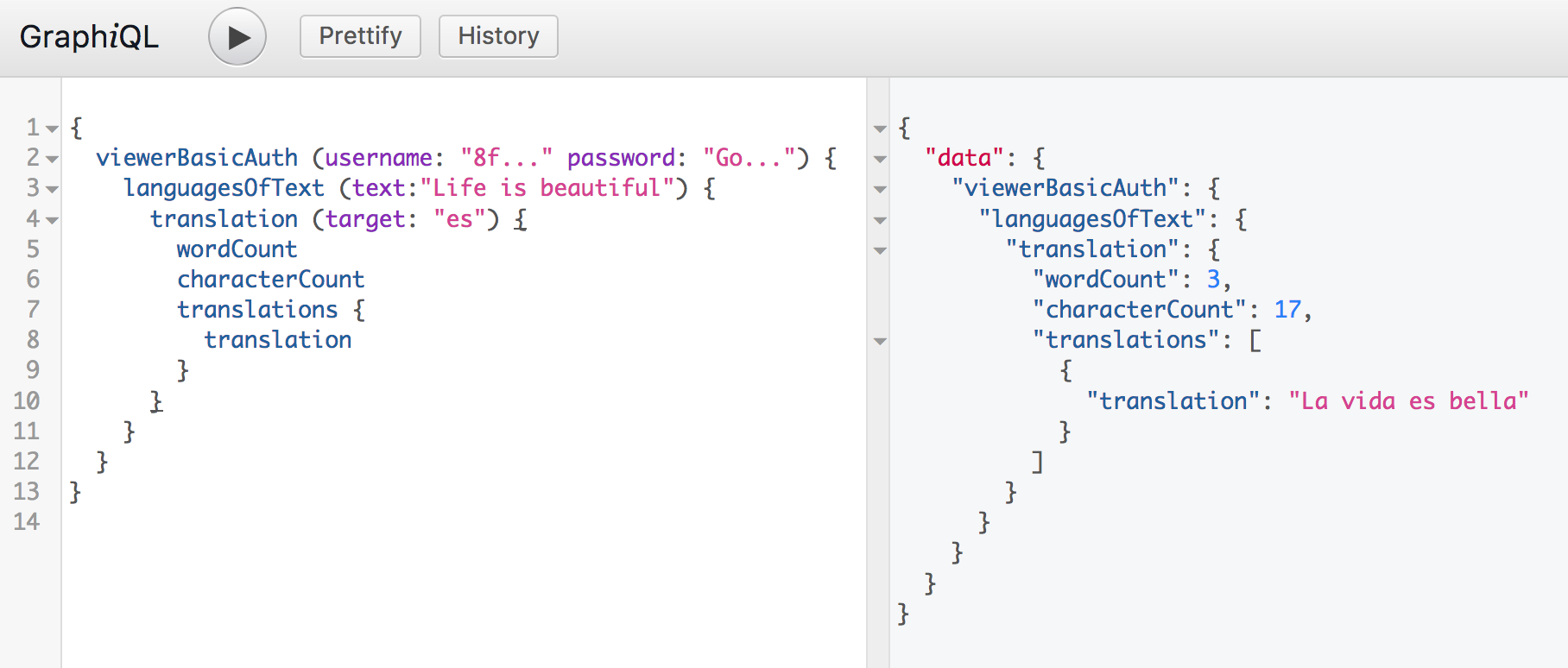}
\caption{GraphQL query and results to detect the language of and translate a text.}
\label{fig:graphiql}
\end{figure}

The resulting wrapper is used as shown in Figure~\ref{fig:graphiql}. A query on the left hand side indicates basic authentication, a string to be translated, and Spanish, \texttt{"es"}, as the target language selection. The right hand side of the figure shows the results of this query, which the GraphQL wrapper produced by composing two (authenticated) requests: one to identify the languages of the given text, and one to translate the text from the identified language to Spanish.
The described changes, neither require changes to the target REST(-like) API, nor they adapt the OAS in a way that breaks its functionality in other contexts.

\section{Related Work}
\label{sec:related_work}
We are not aware of any scientific work that, as we do here, addresses the problem of wrapping REST(-like) APIs using GraphQL. We are only aware of one existing open source tool with the same goal, Swagger2GraphQL\footnote{\url{https://github.com/yarax/swagger-to-graphql}}, which we compare against in the quantitative evaluation in Section~\ref{sec:evaluation}.
In addition, various open source tools generate GraphQL schemas based on given database schemas, for example \emph{PostGraphile} for PostgreSQL\footnote{\url{https://www.graphile.org/postgraphile}}, \emph{tuql} for sqlite\footnote{\url{https://github.com/bradleyboy/tuql}}, or \emph{sql-to-graphql} for SQL databases in general\footnote{\url{https://github.com/rexxars/sql-to-graphql}}.
Given the lack of similar scientific work, in the following, we discuss work that either complements ours or follows a similar goal in a broader sense.

Previous work has attempted to automate the generation of REST(-like) API specifications, like OAS. One approach is to infer specifications from observed dynamic traces of a web server hosting an API~\cite{suter2015inferring}. Another approach is infer specifications from proxied HTTP requests, which can be done by parties other than the provider of an API~\cite{Sohan:2015dp}. While these works do not have the same goal of generating GraphQL wrappers, they help making specifications more broadly available and thus complement the here presented work.

Another branch of previous work addresses the generation of REST(-like) API implementations from different types of specifications. For one, modeling approaches have been proposed that, using methods of model-driven engineering, produce REST(-like) API implementations~\cite{Laitkorpi:2009,Zolotas2017}, and in some cases additionally client code~\cite{Ed-douibi:2016}.
To generate REST(-like) API clients, which a GraphQL wrapper ultimately is as well, related work has proposed to rely on domain-specific languages, which also allow to generate clients that compose requests across APIs~\cite{Maximilien:2007}. The here presented work addresses composition of requests for a single API relying on link definitions in an OAS (see Section~\ref{sec:translation_links}), but could be extended to compose requests across multiple APIs in the future. Other related work discusses advantages of using meta-programming vs. meta-modeling for generating API clients~\cite{Scheidgen:2016}, or the creation of chat bots based on API specifications~\cite{Vaziri:2017}.

In summary, scientific work has not yet addressed the generation of GraphQL wrappers for REST(-like) APIs.

\section{Discussion and Conclusion}
\label{sec:conclusion}
Within this work we presented means to generate GraphQL wrappers for existing REST(-like) APIs based on machine-readable specifications of those APIs (e.g., OAS). We outlined the resulting challenges, like de-duplicating (Input) Object types, sanitizing type, arguments, and field names, dealing with authentication, and enabling nested queries using link definitions.

In experiments with our proof-of-concept implementation, \TOOL, and an open source alternative, Swagger2GraphQL, we assessed how well APIs with publicly available OAS can be wrapped by GraphQL. We find that many OAS, while syntactically correct, have missing or ambiguous information that hinders a complete or exact wrapping. Many of these issues can easily be fixed, though, by completing or correcting the OAS. 

Beyond determining whether REST(-like) APIs can at all be wrapped by GraphQL, the question arises of how usable the generated wrappers are.
For one, GraphQL interfaces should arguably enable nested queries. We describe how nesting can be enabled based on link definitions in an OAS (see Section~\ref{sec:translation_links}). The majority of publicly available OAS, like the ones in APIs.guru, are using OAS version 2.0, though, which lacks support for link definitions. In our use case, we exemplify adding link definitions to an OAS and the effect they have for the resulting GraphQL wrapper (see Section~\ref{sec:usecase}).
Another aspect regarding usability is the question how ``natural'' the GraphQL interface feels. In an idiomatic GraphQL query interface, for example, field names should refer to names of types (e.g., \texttt{User}) rather than the name of an operation (e.g., \texttt{getUser}). \TOOL attempts to adhere to these practices by relying on references and schema object titles to name types, arguments, and fields if possible (see Section~\ref{sec:translation_data_dedup}). We demonstrate how small changes to an OAS can help achieve this goal in the use case (see Section~\ref{sec:usecase}). We consider a more extensive evaluation of the usability of GraphQL wrappers to be future work.
Another aspect of usability concerns the application and evolution of created wrappers. \TOOL builds wrappers in memory, and relies on other tools or libraries to use them. Another option to explore in the future is for \TOOL to output the whole source code of a GraphQL interface, or at least the GraphQL type definitions. On the one hand, generated source code allows developers to customize generated GraphQL interfaces. On the other hand, customizations typically make it hard to re-generate the interface later on, for example as the input API specification evolves.

A major thread of future work for us is to further improve our proof-of-concept implementation.
For one, \TOOL currently lacks means to support pagination, if it is not handled by the REST(-like) target API itself. Especially for nested queries, pagination is important to avoid causing excessive numbers of requests to the REST(-like) API, thus wasting API quotas, hitting rate limits, or even inducing cost.
We would further like to add a caching layer to \TOOL, which can again delimit the strain on the REST(-like) target API.

Finally, two further directions for future work include creating GraphQL wrappers across more than one API, as well as enabling developers to build upon generated GraphQL wrappers and modify them to their liking.

%
%
%
\bibliographystyle{splncs04}
%
\bibliography{bibliography}
\end{document}